\documentclass[aps,prl,groupedaddress,showpacs,amsmath,,reprint,amssymb,floatfix]{revtex4-1}
\usepackage{graphicx}
\usepackage{bm}
\usepackage{color}
\usepackage{epstopdf}
\usepackage{mciteplus}
\usepackage{amsmath, amsthm, amssymb}
\usepackage{latexsym}

\newcommand{\el}{\bm{l}}
\newcommand{\elb}{\bm{l}_B}

\newcommand{\st}{\bm{\varepsilon}}
\newcommand{\Hna}{\bm{H_0}}
\newcommand{\he}{$^3$He}
\newcommand{\hefour}{$^4$He}
\newcommand{\she}{superfluid $^3$He}

\newcommand{\et}{{\it et al.}}

\begin{document}

\title{Orbital-Flop Transition of Angular Momentum in a Topological Superfluid}

\author{A.M. Zimmerman}
\email[]{andrewzimmerman2016@u.northwestern.edu}
\author{J.I.A. Li}
\altaffiliation{Present Address: Department of Physics, Brown University, Providence, RI 02912, USA}
\author{M.D. Nguyen}
\author{W.P. Halperin}
\email[]{w-halperin@northwestern.edu}
\affiliation{Northwestern University, Evanston, IL 60208, USA}

\date{\today}

\begin{abstract}
The direction of the orbital angular momentum of the $B$-phase of superfluid \he\ can be controlled by engineering the anisotropy of the silica aerogel framework within which it is imbibed. In this work, we report our discovery of an unusual and abrupt `orbital-flop' transition of the superfluid angular momentum  between orientations perpendicular and parallel to the anisotropy axis. The transition has no hysteresis, warming or cooling, as  expected for a continuous thermodynamic transition, and is not the result of a competition between strain and magnetic field. This demonstrates the  spontaneous reorientation of the order parameter of an unconventional BCS condensate. 
\end{abstract}

\maketitle

\begin{figure*}
\centerline{\includegraphics[width=0.9\textwidth]{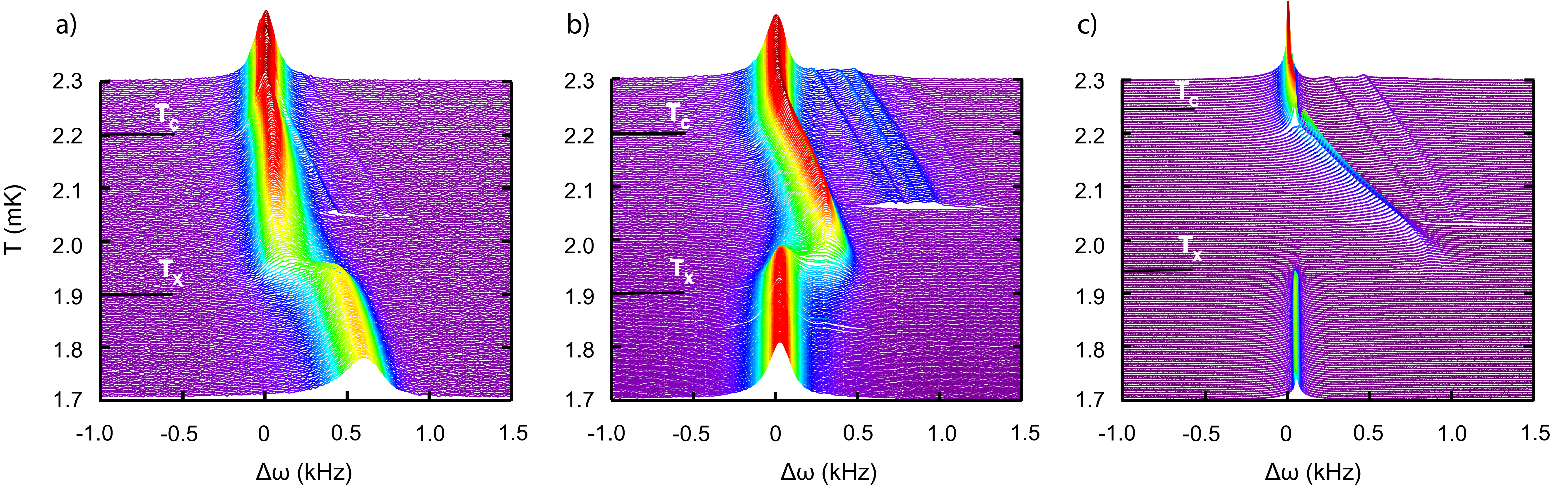}}
\caption{\label{fig:water}(Color online). Temperature dependence of the NMR spectra at small tip angle $\beta\sim 10^\circ$ showing an orbital reorientation transition at $T_x$. Aerogel samples with $\st \sim-20\,\%$ strain at $\hat{H}_0 = 0.096$\,T;  a) $\st \sim -23\,\% $, $\st\perp\Hna$, 26 bar; b)  $\st \sim -19\,\%$, $\st\parallel\Hna$, 26 bar; both a) and b) without $^4$He pre-plating; and c) the same sample as b) but with $^4$He pre-plating at 27 bar. The superfluid transition temperatures are $T_c=$ 2.19, 2.19 and 2.24 mK for a) through c), respectively. The ripples at high frequency shift are from small bulk superfluid components in the $A$-phase.}
\end{figure*}

Unconventional superconductivity and superfluidity break  symmetries of the normal  Fermi liquid state beyond gauge symmetry.  In the case of superfluid \he, the $p$-wave,  spin-triplet condensate has angular momentum $L=1$ and spin angular momentum  $S=1$, and is described by a $3 \times 3$ complex matrix dependent on spin and orbital space vectors.\cite{Leg.75,Vol.90} The spin degrees of freedom couple to the magnetic field, while the orbital degrees of freedom  interact strongly with physical boundaries. Spin and orbit are then coupled through the nuclear dipole-dipole interaction. Here, we investigate the effects of impurities on \she-B relevant to understanding disorder in quantum condensed states with a vector order parameter. In the present work, these  impurities are in the form of high-porosity silica aerogel which consists of microscopic particles, $\sim 3$\,nm diameter, in a limited fractal distribution.\cite{Pol.08} In the presence of anisotropic disorder from strained aerogel, we have discovered a transition in orientation of the orbital angular momentum  in \she.  We detect this transition through the orientation of the angular momentum axis relative to the magnetic field from nuclear magnetic resonance (NMR) spectra  shown in Fig.\,\ref{fig:water}.

Shortly after the  discovery of superfluidity of \he\ imbibed in silica aerogel,\cite{Por.95,Spr.95} the experimental observations of this new class of coherent superfluid states were accounted for with a Ginzburg-Landau theory.\cite{Thu.98, Sha.01} The principal parameters of the model are the Fermi quasiparticle  mean-free-path, $\lambda$, and a silica particle-particle correlation length, $\xi_a$.  An important prediction was that anisotropic (isotropic) Fermi quasiparticle scattering would tend to favor the stability of anisotropic (isotropic) superfluid states. This provoked development and characterization of aerogel samples with uniform, global anisotropy, as well as 'isotropic aerogels'  with no preferred direction on length scales greater than a few microns.\cite{Pol.08} NMR experiments on the superfluid imbibed in isotropic samples were consistent with these predictions.\cite{Pol.11} This was followed by a series of experiments and theoretical works focusing on anisotropic superfluids and anisotropic disorder\cite{Pol.12a,Dmi.10,Li.15} that has recently been summarized.\cite{Hal.18}

\begin{figure}[t!]
\centerline{\includegraphics[width=0.5\textwidth]{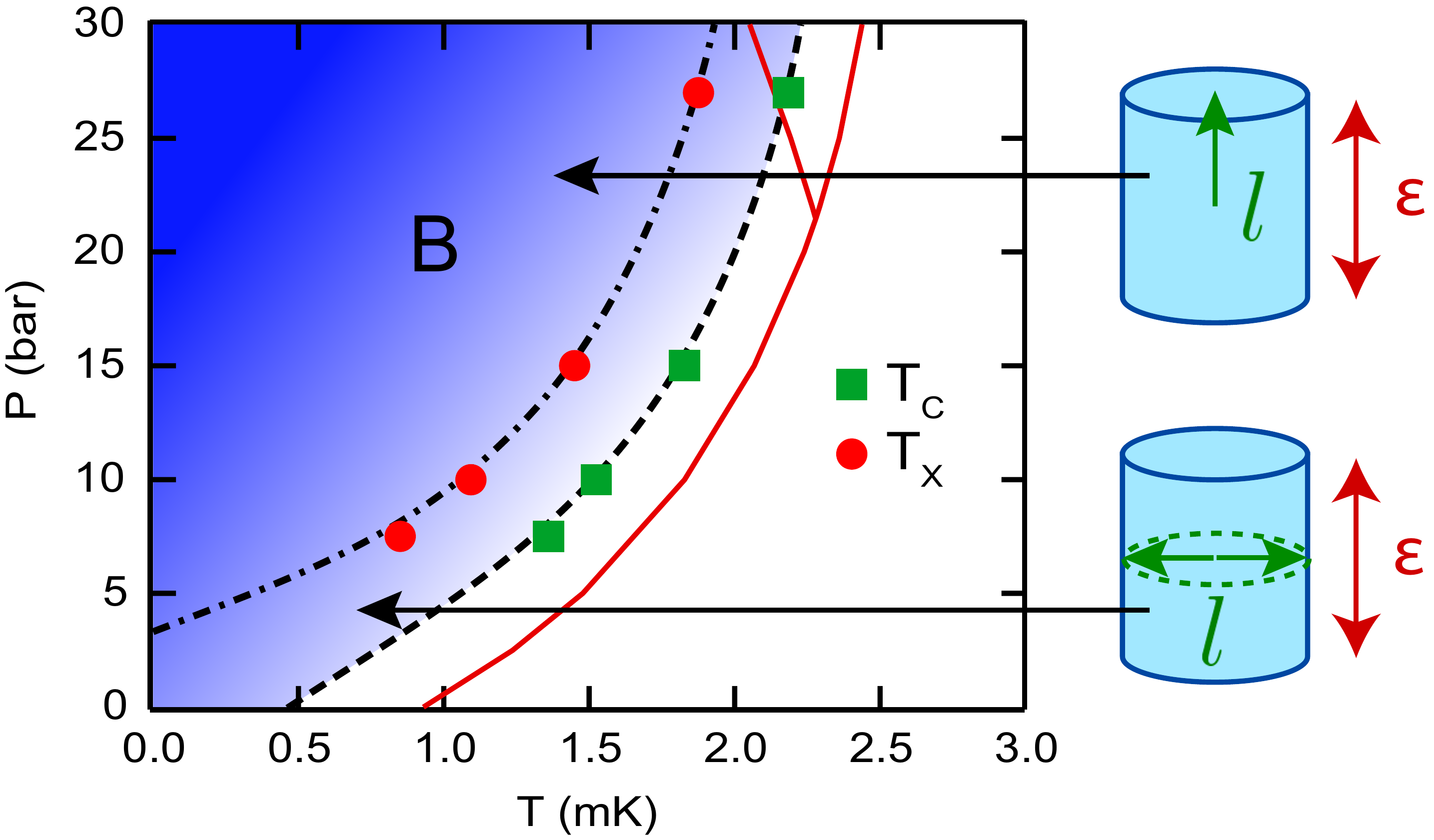}}
\caption{\label{fig:PVT}(Color online) Pressure-temperature phase diagram at low magnetic field, extrapolated to zero, for an aerogel compressed by $\sim20\%$ from an isotropic sample reported upon earlier.\cite{Pol.11} The green squares  correspond to the superfluid phase transition, $T_c$, in aerogel, and the solid red circles to the reorientation transition, $T_x$, without $^4$He pre-plating. The dashed lines are guides to the eye. The solid red lines are the phase diagram of pure superfluid $^3$He. }
\end{figure} 

\begin{figure}[b]
\centerline{\includegraphics[width=0.5\textwidth]{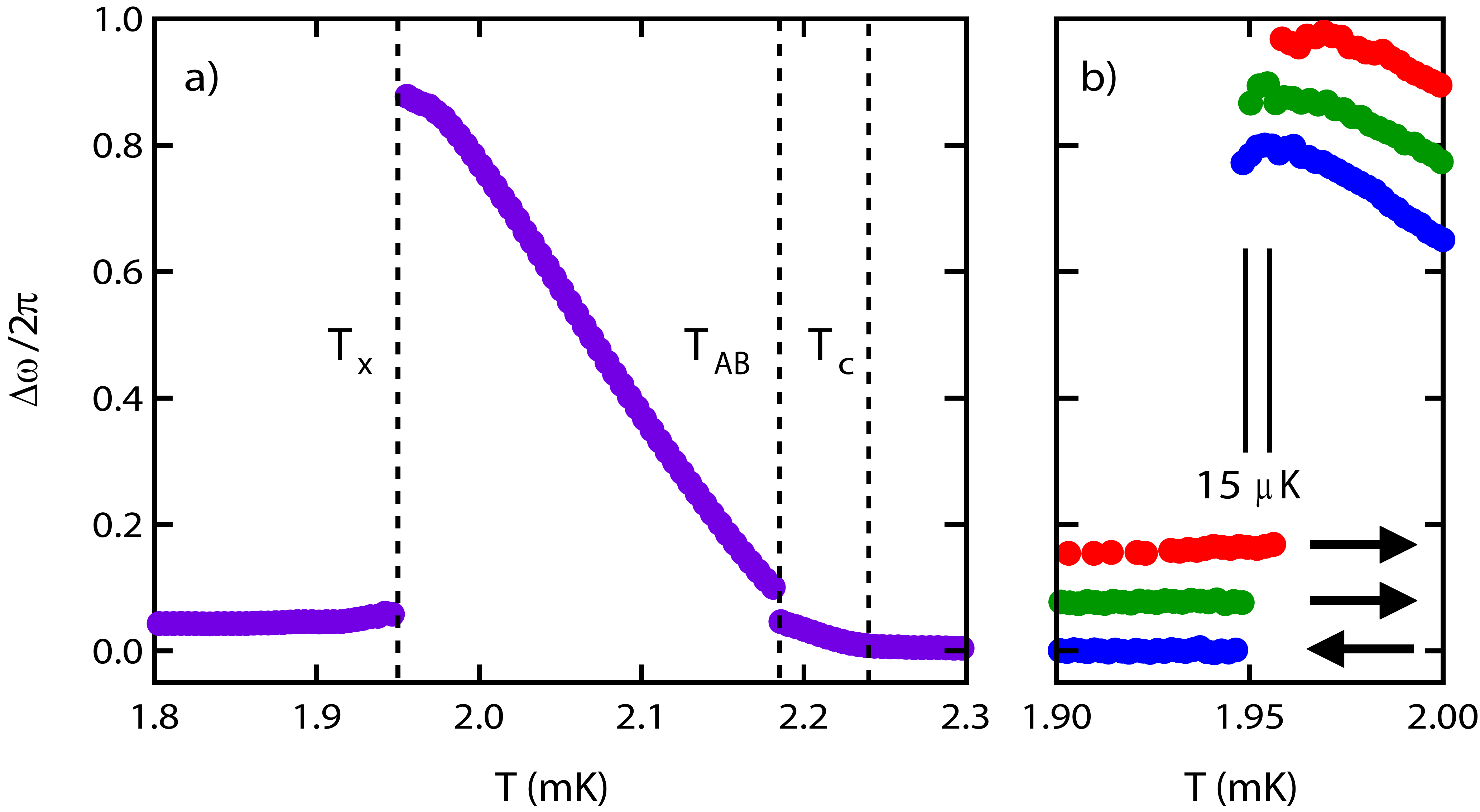}}
\caption{\label{fig:freq}(Color online). a) NMR frequency shift, $\Delta\omega$, as a function of temperature with $\st\parallel\Hna$ and $^4$He pre-plating at $\hat{H}_0 =0.096$\,T and $P=27$\,bar. This data corresponds to the spectra in Fig.\,\ref{fig:water}c). The large $\Delta\omega$ for $T>T_x$ indicates $\elb\perp\Hna$ and the small $\Delta\omega$ at low temperatures, $\elb\parallel\Hna$. There is a small window of the superfluid A-phase close to $T_c$. b) (Traces are offset vertically for clarity.) Tracking $\Delta\omega$ back and forth across $T_x$ in directions indicated by arrows, shows no hysteresis within the reproducibility of the measurement of temperature, $\sim15\,\mu$K. }
\end{figure} 

Although the order parameter amplitude for the pure superfluid $B$-phase is completely isotropic, in the presence of an external field that breaks this perfect orbital rotation symmetry (sample boundaries, phase boundaries, superfluid flow, magnetic fields, and anisotropic disorder from impurities on length scales much larger than the superfluid coherence length) its orbital degrees of freedom are constrained, and can be characterized by an orbital angular momentum axis, $\elb$.\cite{Hak.89}  This axis is characterized by a large bending energy, resulting in order parameter textures.\cite{Vol.90}  In particular, spatial constraints from sample cell walls and external surfaces produce a ``flare-out" texture for \he-B in its pure form,\cite{Hak.89} as well as in isotropic aerogel.\cite{Pol.11}  For negatively strained  aerogel  the situation is very different.

Contrary to expectation, a distorted version of the $B$-phase becomes progressively more stable with increasing negative strain, effectively competing with the $A$-phase, persisting up to $T_c$ and  extending to non-zero magnetic fields.\cite{Li.14b} In the present work, we have exploited this fact to explore the orientation effects of anisotropic disorder on the orbital components of the $B$-phase order parameter over our full temperature range up to $T_c$, Fig.\,\ref{fig:water}, and as a function of pressure, Fig.\,\ref{fig:PVT}. We find that the orientation of $\elb$ is  controlled by the strain axis, $\st$, as was first shown in NMR measurements by Li \et\cite{Li.14a}, Fig.\,\ref{fig:water}\,a), b) at 26 bar. Here, we have carried out a thorough  investigation of the orientation of $\elb$, including increasing our NMR spectral resolution by almost two orders of magnitude, Fig.\,\ref{fig:water}\,c), by  eliminating two monolayers of paramagnetic  solid  \he\ adsorbed on the aerogel surface with the substitution of solid \hefour, so-called pre-plating. When present, adsorbed solid \he\ is in fast exchange with the bulk liquid, decreasing the frequency shift for the entire sample by the ratio of liquid-to-solid magnetization as well as increasing the spectral linewidth.\cite{Spr.95,Col.09} With our increased frequency resolution, it appears that there is a very well-defined temperature $T_x$ above which the order parameter is oriented with $\elb\perp\st$, while below, it reorients with $\elb\parallel\st$.  This transition is sharp and reversible, while the magnetic susceptibility is continuous with temperature throughout, as is characteristic of the pure $B$-phase.  It appears that we can flip the angular momentum orientation $90^{\circ}$ by changing the temperature by as little as $15\,\mu$K, Fig.\,\ref{fig:freq}, or by what should be equivalent, changing the pressure, Fig.\,\ref{fig:PVT}.

\begin{figure}[t]
\centerline{\includegraphics[width=0.45\textwidth]{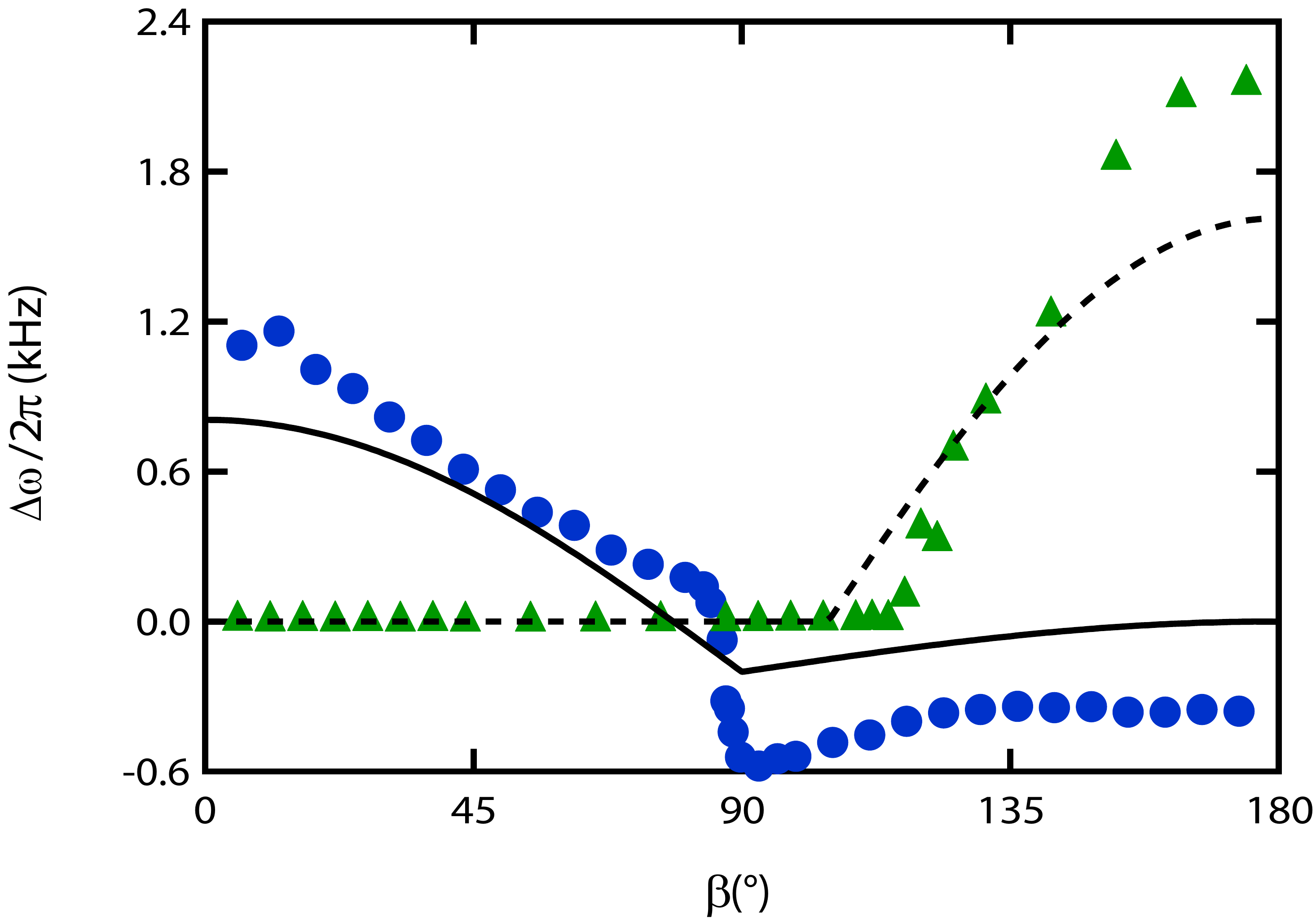}}
\caption{\label{fig:tip}(Color online).  Typical tipping angle dependence of the frequency shift of data above $T_x$ at $T=2.03$ mK (blue circles) for the sample in Fig.\,\ref{fig:water}\,c), and below $T_x$ at $T=1.76$ mK, scaled to 2.03 mK. Measurements were taken with $\st\parallel\Hna$ and $^4$He pre-plating at 0.05 T. The solid (dashed) curve shows the theoretical dependence for $\Delta\omega_{\perp}$ ($\Delta\omega_{\parallel}$) scaled to the frequency shift of isotropic aerogel.\cite{Pol.11} The larger magnitude frequency shift with negative strain, as compared with the isotropic case, results from strain-induced  distortion of the $B$-phase.\cite{Li.15}  }
\end{figure}

Pulsed $^3$He NMR measurements were conducted with a magnet having a homogeneity of $\approx 2$\, ppm over the sample volume and magnetic fields from 50 to 200 mT.  Temperature was measured with $^{195}$Pt NMR and melting curve thermometry, both with reference to the known NMR signatures of the superfluid transitions from small amounts of bulk \he\ in the sample cell, evident in Fig.\,\ref{fig:water}.
Cylindrical silica aerogel samples, 4\,mm diameter $\times$ 5\,mm length, were grown in glass tubes with nominal 98\% porosity.  Following a one-step procedure with rapid supercritical extraction, we used optical birefringence\cite{Pol.08} to obtain a quantitative measure of its anisotropy.\cite{Li.15}  After  investigation of \she\ imbibed in one isotropic sample,\cite{Pol.11, Li.13} we compressed it by $\sim20\%$ to provide the data in Fig.\,\ref{fig:water}\,b),\,c), and also in Fig.\,\ref{fig:PVT}-\ref{fig:tip}.  The early work with this sample was performed without \hefour\ pre-plating, Fig.\,\ref{fig:water}\,b), as were experiments with other similarly prepared samples, such as Fig.\,\ref{1}\,a). 

The order parameter orientation in the $B$-phase was determined from the peak frequency of the NMR spectra.\cite{Vol.90}  The coupling of spin and orbital degrees of freedom results in a frequency shift, $\Delta\omega$, away from the normal-state precession at the Larmor frequency, $\omega_L$.\cite{Leg.73,Zim.18} This  shift is dependent on both the magnitude and orientation of the order parameter, as well as the amplitude of the applied radio frequency pulsed excitation, which tips the nuclear magnetization by an angle $\beta$ with respect to the static magnetic field, $\Hna$.  For \he-B there are two experimentally realized limits which illustrate the effect of the orientation of $\elb$ on the NMR response to this excitation.  If $\elb\parallel\Hna$, the frequency shift is given as a function of $\mathbf{\beta}$ as,\cite{Bri.75, Cor.78}
\begin{eqnarray}
      \Delta\omega_{\parallel}&\approx&0,\hspace{110pt}\beta < 104^{\circ}\label{1}\\
      \Delta\omega_{\parallel}&=&-\frac{4}{15}\frac{\Omega_{B}^{2}}{\omega_{L}}(1+4\cos\beta).\hspace{16pt}\beta > 104^{\circ}, \label{2}
 \end{eqnarray}
\noindent
where $\Omega_{B}$ is the longitudinal resonance frequency of the $B$-phase and is proportional to the amplitude of the order parameter.\cite{Leg.75} When $ \beta$ is small, this results in near-zero measured frequency shift, with the onset of larger shifts  for tipping angles greater than the Leggett angle, $104^\circ$. This is the expected behavior of the bulk superfluid in a uniform magnetic field, in the absence of any external constraints.
The other limit, $\elb\perp\Hna$, is not the minimum of the dipole energy, and so requires some external constraint imposed on a higher energy scale to orient the angular momentum.  If $\elb\perp\Hna$, the resulting frequency shift is,\cite{Yur.93b,Dmi.99}
\begin{eqnarray}
     \Delta\omega_{\perp}&=&\frac{\Omega_{B}^{2}}{2\omega_{L}}(\cos\beta-\frac{1}{5}),\hspace{30pt}\beta < 90^{\circ},\label{3}\\
      \Delta\omega_{\perp}&=&-\frac{\Omega_{B}^{2}}{10\omega_{L}}(1+\cos\beta).\,\hspace{20pt}\beta > 90^{\circ}. \label{4}
\end{eqnarray}
\noindent
In this case, there is a large positive frequency shift measured at small tip angles. This type of behavior has been reported for \she-B in both its pure form\cite{Vol.90} and in isotropic silica aerogel,\cite{Pol.11} where $\elb$ is constrained to be normal to a surface.  These two limits are displayed in Fig.\,\ref{fig:tip}, where the dashed black curve corresponds to $\Delta\omega_{\parallel}$, Eq.'s\,\ref{1},\,\ref{2}, and the solid curve for $\Delta\omega_{\perp}$ from Eq.'s\,\ref{3},\,\ref{4}.

The configuration $\elb\parallel\Hna$ corresponds to the behavior we observe for $T>T_x$ with $\st\perp\Hna$ (Fig.\,\ref{fig:water}\,a) and for $T<T_x$ when $\st\parallel\Hna$ (Fig.\,\ref{fig:water}\,b),\,c). For the complementary ranges of temperature in these two cases, the spectra indicate that $\elb\perp\Hna$.  The tip angle dependence of the frequency shift is shown for both regions in Fig.\,\ref{fig:tip}, showing qualitative agreement.  This is one example from among many measurements, confirming our identification of these orientations of the angular momentum axis relative to the magnetic field.  The NMR linewidths of the spectra in Fig.\,\ref{fig:water}\,a),\,b) are approximately the same as the normal state, indicating that the entire sample has a uniform orientation of $\elb$.  For the higher resolution data, Fig.\,\ref{fig:water}\,c), the linewidth increases slightly at high temperatures, $T>T_x$,  possibly indicating a small deviation from a uniform distribution, since this configuration is not a minimum of the dipole energy. We note that the linewidth is always smaller than that associated with the textures in pure \he-B or the $B$-phase of isotropic aerogel. The temperature at which the transition occurs, $T_x$,  is very well-defined and non-hysteretic within experimental error, Fig.\,\ref{fig:freq}b).  

\begin{figure}
\centerline{\includegraphics[width=0.45\textwidth]{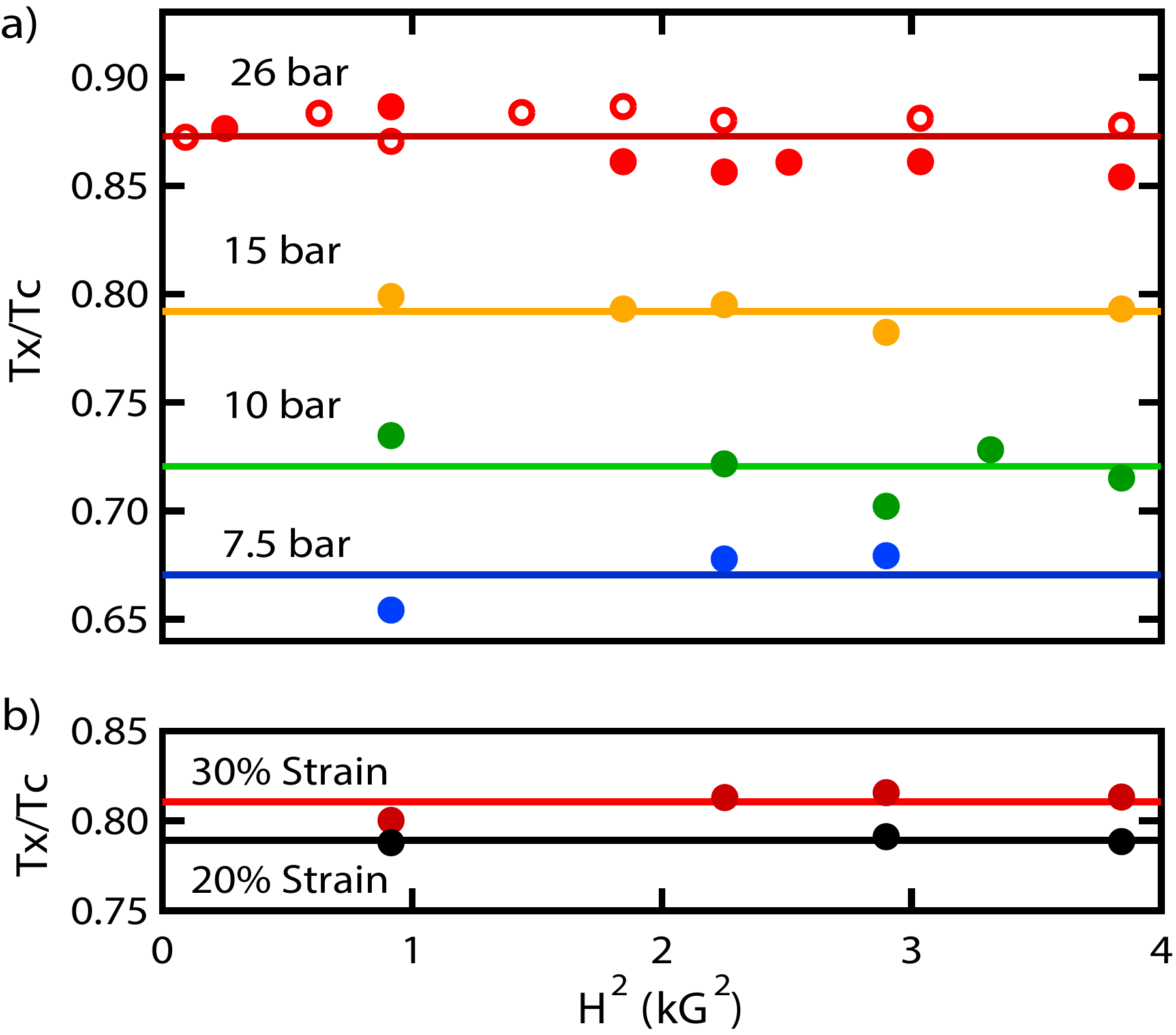}}%
\caption{\label{fig:field}(Color online) a) Magnetic field independence of $T_x/T_{c}$ for different pressures without pre-plating. At $P=26$\,bar with $\sim-19\%$ strain, $\st\parallel\Hna$ (open circles), and with $\sim-23\%$, $\st\perp\Hna$ (closed circles). b) Strain dependence is very small between $\sim-20\%$ and $\sim-30\%$ at 27 bar with $\st\perp\Hna$. }
\end{figure}

\begin{figure}[t]
\centerline{\includegraphics[height=0.25\textheight]{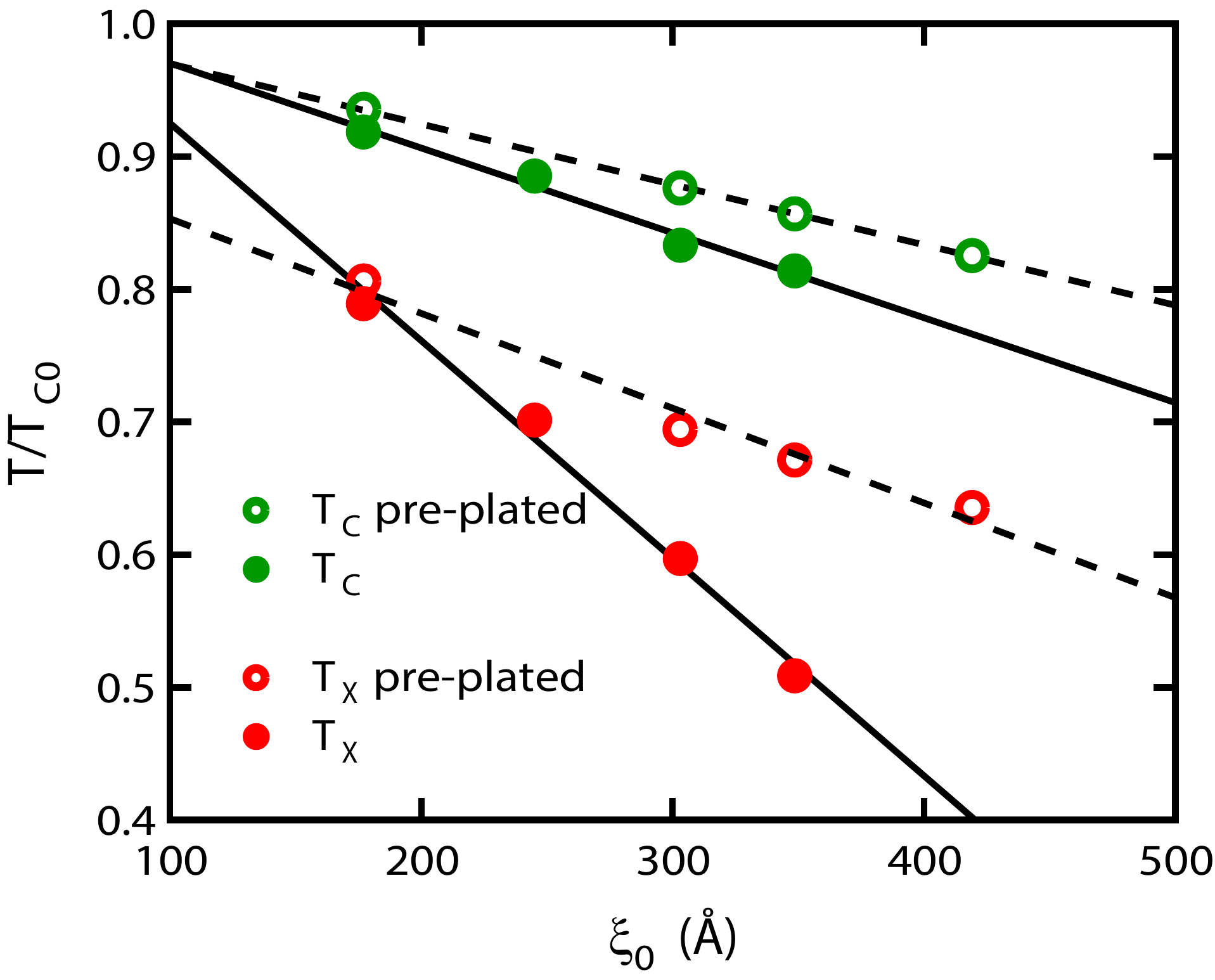}}
\caption{\label{fig:coherence} (Color online) Transition temperatures for $T_{c}$ and $T_x$ scaled to the pure transition temperature as a function of the pure superfluid coherence length. Closed (open) circles indicate measurements without (with) $^4$He pre-plating. The linear dependence suggests that both transitions are driven by impurity and can be characterized by a length scale in the aerogel. Without pre-plating, the extracted length scales are $\lambda = 385$ nm and $\lambda_x = 150$ nm. With pre-plating, they are $\lambda = 550$ nm and $\lambda_x = 347$ nm. For all data, we find $\xi_a \sim 3 - 6$ nm.  These should be compared with  estimates for the the length scales for 98\% porosity isotropic aerogel, $\lambda \approx 150$\,nm and $\xi_a \approx 40$\,nm.\cite{Hal.08}} 
\end{figure}

This behavior has been investigated in samples with four values of strain: $0\%$, $\sim-12\%$, $\sim-20\%$, and $\sim-30\%$. We find that $T_x$  is approximately independent of strain between $-20\%$ and $-30\%$, Fig.\,\ref{fig:field}\,b). The transition is not present in  isotropic samples, or for small strain $\sim-12\%$, indicating that there is an anisotropy threshold for the transition to exist above which it is independent.  Similarly, $T_x$ is independent of the magnitude of magnetic field and its orientation, Fig.\,\ref{fig:field}, indicating that the reorientation of the order parameter cannot be the result of competition between the strain and magnetic field. $T_x$ is modified only by aerogel surface conditions and pressure, Fig.\,\ref{fig:PVT} and \ref{fig:coherence},  demonstrating that the transition is driven by the impurity.  From the known relative orientations of the strain axis and magnetic field  we conclude that for  $T>T_x$  the angular momentum is perpendicular to the strain axis, $\elb\perp\st$, and that for $T<T_x$ it reorients to be $\elb\parallel\st$.  We do not cover a sufficiently large range of magnetic field to stabilize the $A$ phase at temperatures below $T_x$; however, we note that above this temperature the angular momentum in the $A$-phase is oriented  with $\el\perp\st$, the same as  that of the $B$-phase at the same temperatures, but at lower magnetic fields.\cite{Li.14b}

Various physical systems with vector order parameters exhibit transitions involving their reorientation, most notably the Fr\'eedericksz transition in liquid crystals\cite{deG.95}, important in flat screen displays, and the spin-flop transition in antiferromagnets\cite{DeJ.01}, as well as multiferroics\cite{Son.09}, which have a variety of electronic switching applications. Additionally, a reorientation Fr\'eedericksz-type transition of $\el$ in \she-A was reported, a balance between magnetic field and superfluid flow from rotation.\cite{Yam.08}  All of these transitions are characterized by a competition between two orienting fields. The transition at $T_x$ is quite unique, appearing spontaneously as a function of temperature.

The interaction of a superconductor with disorder is determined by the superconducting coherence length, $\xi_0$. To describe  a conventional, $L=0$ superconductor in the presence of isotropic disorder, Abrikosov and Gorkov\cite{Abr.62} developed a theory that expressed the superconducting transition in terms of a pair-breaking parameter, $x$, given by the ratio of the pure sample coherence length to the quasiparticle mean-free-path associated with quasiparticle scattering from magnetic impurities, $x = \xi_0/\lambda$. The theory was extended to isotropic impurity in unconventional superconductors, by Larkin\cite{Lar.65}, where both magnetic and nonmagnetic scattering play a role, and later specialized to \she.\cite{Thu.98,Fom.08} 
In the limit of small suppression of $T_c$ relative to $T_{c0}$, the transition temperature  is linear in $x$, 
\begin{equation}\label{5}
\frac{T_{c}}{T_{c0}}=1-\frac{\pi^2}{4}\frac{\xi_0}{\lambda},
\end{equation}
where $T_{c0}$ is the impurity-free transition temperature.

This expression must be modified due to the fractal structure of aerogel by taking correlations into account. Sauls and Sharma\cite{Sau.03} redefined the pair-breaking parameter to allow for aerogel silica particle-particle correlations with a correlation length $\xi_a$ by introducing an effective pair-breaking parameter $\tilde{x}=x/(1+\xi_a^2/(\lambda^2x))$. A similar result was derived by Fomin by different methods.\cite{Fom.08}  In the clean limit as $T_c\rightarrow T_{c0}$,
\begin{equation}\label{6}
\frac{T_{c}}{T_{c0}}=1-\frac{\pi^2}{4}\frac{\xi_0}{\lambda} + \frac{\pi^2}{4}\left(\frac{\xi_a}{\lambda}\right)^2. 
\end{equation}

In Fig.\,\ref{fig:coherence}, we show that both $T_{c}$ and $T_x$ vary linearly with $\xi_0$ for \she\ in aerogel, suggesting that both are described by quasiparticle scattering from impurities. The different slopes for $T_c$ and $T_x$  indicate that a new length scale, $\lambda_x$, must be introduced to describe $T_x$, taking the place of $\lambda$ in Eq.\,\ref{6}. After pre-plating, different values of $\lambda$ and $\lambda_x$ are extracted due to changes in the scattering conditions. The ratio $\lambda/\lambda_x=1.6$ with pre-plating and 2.6 otherwise. It is much larger than the aerogel anisotropy of $\sim 20\%$. This, together with the strain independence in Fig.\,\ref{fig:field} appears to rule out the simple identification of $\lambda$ and $\lambda_x$ as the mean free path perpendicular and parallel to strain. However, it is clear that $\lambda_x$ is a signature of the strain-induced broken rotational symmetry from the aerogel structure. Determining the exact nature of this new length scale and the mechanism for the angular momentum reorientation will require further investigation.  

In summary, we have discovered a unique `orbital flop' transition in \she-B in negatively strained silica aerogel. It appears as a spontaneous reorientation of the vector order parameter relative to the direction of strain  in the aerogel structure. It is a magnetic field-independent, order parameter transition that varies with pressure and aerogel surface conditions and is ubiquitous in samples compressed beyond a threshold. 

We are grateful to J. A. Sauls, J. J. Wiman, J. W. Scott, V. V. Dmitriev, and G. E. Volovik for helpful discussion, and support from the National Science Foundation (Grant No. DMR-1602542).


\begin{thebibliography}{32}%
\makeatletter
\providecommand \@ifxundefined [1]{%
 \@ifx{#1\undefined}
}%
\providecommand \@ifnum [1]{%
 \ifnum #1\expandafter \@firstoftwo
 \else \expandafter \@secondoftwo
 \fi
}%
\providecommand \@ifx [1]{%
 \ifx #1\expandafter \@firstoftwo
 \else \expandafter \@secondoftwo
 \fi
}%
\providecommand \natexlab [1]{#1}%
\providecommand \enquote  [1]{``#1''}%
\providecommand \bibnamefont  [1]{#1}%
\providecommand \bibfnamefont [1]{#1}%
\providecommand \citenamefont [1]{#1}%
\providecommand \href@noop [0]{\@secondoftwo}%
\providecommand \href [0]{\begingroup \@sanitize@url \@href}%
\providecommand \@href[1]{\@@startlink{#1}\@@href}%
\providecommand \@@href[1]{\endgroup#1\@@endlink}%
\providecommand \@sanitize@url [0]{\catcode `\\12\catcode `\$12\catcode
  `\&12\catcode `\#12\catcode `\^12\catcode `\_12\catcode `\%12\relax}%
\providecommand \@@startlink[1]{}%
\providecommand \@@endlink[0]{}%
\providecommand \url  [0]{\begingroup\@sanitize@url \@url }%
\providecommand \@url [1]{\endgroup\@href {#1}{\urlprefix }}%
\providecommand \urlprefix  [0]{URL }%
\providecommand \Eprint [0]{\href }%
\providecommand \doibase [0]{http://dx.doi.org/}%
\providecommand \selectlanguage [0]{\@gobble}%
\providecommand \bibinfo  [0]{\@secondoftwo}%
\providecommand \bibfield  [0]{\@secondoftwo}%
\providecommand \translation [1]{[#1]}%
\providecommand \BibitemOpen [0]{}%
\providecommand \bibitemStop [0]{}%
\providecommand \bibitemNoStop [0]{.\EOS\space}%
\providecommand \EOS [0]{\spacefactor3000\relax}%
\providecommand \BibitemShut  [1]{\csname bibitem#1\endcsname}%
\let\auto@bib@innerbib\@empty
\bibitem [{\citenamefont {Leggett}(1975)}]{Leg.75}%
  \BibitemOpen
  \bibfield  {author} {\bibinfo {author} {\bibfnamefont {A.~J.}\ \bibnamefont
  {Leggett}},\ }\href@noop {} {\bibfield  {journal} {\bibinfo  {journal} {Rev.
  Mod. Phys.}\ }\textbf {\bibinfo {volume} {47}},\ \bibinfo {pages} {331}
  (\bibinfo {year} {1975})}\BibitemShut {NoStop}%
\bibitem [{\citenamefont {Vollhardt}\ and\ \citenamefont
  {W{\"o}lfle}(1990)}]{Vol.90}%
  \BibitemOpen
  \bibfield  {author} {\bibinfo {author} {\bibfnamefont {D.}~\bibnamefont
  {Vollhardt}}\ and\ \bibinfo {author} {\bibfnamefont {P.}~\bibnamefont
  {W{\"o}lfle}},\ }\href@noop {} {\emph {\bibinfo {title} {The Superfluid
  Phases of Helium 3}}}\ (\bibinfo  {publisher} {Taylor and Francis},\ \bibinfo
  {year} {1990})\BibitemShut {NoStop}%
\bibitem [{\citenamefont {Pollanen}\ \emph {et~al.}(2008)\citenamefont
  {Pollanen}, \citenamefont {Shirer}, \citenamefont {Blinstein}, \citenamefont
  {Davis}, \citenamefont {Choi}, \citenamefont {Lippman}, \citenamefont
  {Lurio},\ and\ \citenamefont {Halperin}}]{Pol.08}%
  \BibitemOpen
  \bibfield  {author} {\bibinfo {author} {\bibfnamefont {J.}~\bibnamefont
  {Pollanen}}, \bibinfo {author} {\bibfnamefont {K.~R.}\ \bibnamefont
  {Shirer}}, \bibinfo {author} {\bibfnamefont {S.}~\bibnamefont {Blinstein}},
  \bibinfo {author} {\bibfnamefont {J.~P.}\ \bibnamefont {Davis}}, \bibinfo
  {author} {\bibfnamefont {H.}~\bibnamefont {Choi}}, \bibinfo {author}
  {\bibfnamefont {T.~M.}\ \bibnamefont {Lippman}}, \bibinfo {author}
  {\bibfnamefont {L.~B.}\ \bibnamefont {Lurio}}, \ and\ \bibinfo {author}
  {\bibfnamefont {W.~P.}\ \bibnamefont {Halperin}},\ }\href@noop {} {\bibfield
  {journal} {\bibinfo  {journal} {J. Non-Crystalline Solids}\ }\textbf
  {\bibinfo {volume} {354}},\ \bibinfo {pages} {4668} (\bibinfo {year}
  {2008})}\BibitemShut {NoStop}%
\bibitem [{\citenamefont {Porto}\ and\ \citenamefont {Parpia}(1995)}]{Por.95}%
  \BibitemOpen
  \bibfield  {author} {\bibinfo {author} {\bibfnamefont {J.~V.}\ \bibnamefont
  {Porto}}\ and\ \bibinfo {author} {\bibfnamefont {J.~M.}\ \bibnamefont
  {Parpia}},\ }\href@noop {} {\bibfield  {journal} {\bibinfo  {journal} {Phys.
  Rev. Lett.}\ }\textbf {\bibinfo {volume} {74}},\ \bibinfo {pages} {4667}
  (\bibinfo {year} {1995})}\BibitemShut {NoStop}%
\bibitem [{\citenamefont {Sprague}\ \emph {et~al.}(1995)\citenamefont
  {Sprague}, \citenamefont {Haard}, \citenamefont {Kycia}, \citenamefont
  {Rand}, \citenamefont {Lee}, \citenamefont {Hamot},\ and\ \citenamefont
  {Halperin}}]{Spr.95}%
  \BibitemOpen
  \bibfield  {author} {\bibinfo {author} {\bibfnamefont {D.~T.}\ \bibnamefont
  {Sprague}}, \bibinfo {author} {\bibfnamefont {T.~M.}\ \bibnamefont {Haard}},
  \bibinfo {author} {\bibfnamefont {J.~B.}\ \bibnamefont {Kycia}}, \bibinfo
  {author} {\bibfnamefont {M.~R.}\ \bibnamefont {Rand}}, \bibinfo {author}
  {\bibfnamefont {Y.}~\bibnamefont {Lee}}, \bibinfo {author} {\bibfnamefont
  {P.~J.}\ \bibnamefont {Hamot}}, \ and\ \bibinfo {author} {\bibfnamefont
  {W.~P.}\ \bibnamefont {Halperin}},\ }\href@noop {} {\bibfield  {journal}
  {\bibinfo  {journal} {Phys. Rev. Lett.}\ }\textbf {\bibinfo {volume} {75}},\
  \bibinfo {pages} {661} (\bibinfo {year} {1995})}\BibitemShut {NoStop}%
\bibitem [{\citenamefont {Thuneberg}\ \emph {et~al.}(1998)\citenamefont
  {Thuneberg}, \citenamefont {Yip}, \citenamefont {Fogelstr{\"o}m},\ and\
  \citenamefont {Sauls}}]{Thu.98}%
  \BibitemOpen
  \bibfield  {author} {\bibinfo {author} {\bibfnamefont {E.~V.}\ \bibnamefont
  {Thuneberg}}, \bibinfo {author} {\bibfnamefont {S.~K.}\ \bibnamefont {Yip}},
  \bibinfo {author} {\bibfnamefont {M.}~\bibnamefont {Fogelstr{\"o}m}}, \ and\
  \bibinfo {author} {\bibfnamefont {J.~A.}\ \bibnamefont {Sauls}},\ }\href@noop
  {} {\bibfield  {journal} {\bibinfo  {journal} {Phys. Rev. Lett.}\ }\textbf
  {\bibinfo {volume} {80}},\ \bibinfo {pages} {2861} (\bibinfo {year}
  {1998})}\BibitemShut {NoStop}%
\bibitem [{\citenamefont {Sharma}\ and\ \citenamefont {Sauls}(2001)}]{Sha.01}%
  \BibitemOpen
  \bibfield  {author} {\bibinfo {author} {\bibfnamefont {P.}~\bibnamefont
  {Sharma}}\ and\ \bibinfo {author} {\bibfnamefont {J.~A.}\ \bibnamefont
  {Sauls}},\ }\href@noop {} {\bibfield  {journal} {\bibinfo  {journal} {J. Low
  Temp. Phys.}\ }\textbf {\bibinfo {volume} {125}},\ \bibinfo {pages} {115}
  (\bibinfo {year} {2001})}\BibitemShut {NoStop}%
\bibitem [{\citenamefont {Pollanen}\ \emph {et~al.}(2011)\citenamefont
  {Pollanen}, \citenamefont {Li}, \citenamefont {Collett}, \citenamefont
  {Gannon},\ and\ \citenamefont {Halperin}}]{Pol.11}%
  \BibitemOpen
  \bibfield  {author} {\bibinfo {author} {\bibfnamefont {J.}~\bibnamefont
  {Pollanen}}, \bibinfo {author} {\bibfnamefont {J.~I.~A.}\ \bibnamefont {Li}},
  \bibinfo {author} {\bibfnamefont {C.~A.}\ \bibnamefont {Collett}}, \bibinfo
  {author} {\bibfnamefont {W.~J.}\ \bibnamefont {Gannon}}, \ and\ \bibinfo
  {author} {\bibfnamefont {W.~P.}\ \bibnamefont {Halperin}},\ }\href@noop {}
  {\bibfield  {journal} {\bibinfo  {journal} {Phys. Rev. Lett.}\ }\textbf
  {\bibinfo {volume} {107}},\ \bibinfo {pages} {195301} (\bibinfo {year}
  {2011})}\BibitemShut {NoStop}%
\bibitem [{\citenamefont {Pollanen}\ \emph {et~al.}(2012)\citenamefont
  {Pollanen}, \citenamefont {Li}, \citenamefont {Collett}, \citenamefont
  {Gannon}, \citenamefont {Halperin},\ and\ \citenamefont {Sauls}}]{Pol.12a}%
  \BibitemOpen
  \bibfield  {author} {\bibinfo {author} {\bibfnamefont {J.}~\bibnamefont
  {Pollanen}}, \bibinfo {author} {\bibfnamefont {J.~I.~A.}\ \bibnamefont {Li}},
  \bibinfo {author} {\bibfnamefont {C.~A.}\ \bibnamefont {Collett}}, \bibinfo
  {author} {\bibfnamefont {W.~J.}\ \bibnamefont {Gannon}}, \bibinfo {author}
  {\bibfnamefont {W.~P.}\ \bibnamefont {Halperin}}, \ and\ \bibinfo {author}
  {\bibfnamefont {J.~A.}\ \bibnamefont {Sauls}},\ }\href@noop {} {\bibfield
  {journal} {\bibinfo  {journal} {Nature Phys.}\ }\textbf {\bibinfo {volume}
  {8}},\ \bibinfo {pages} {317} (\bibinfo {year} {2012})}\BibitemShut {NoStop}%
\bibitem [{\citenamefont {Dmitriev}\ \emph {et~al.}(2010)\citenamefont
  {Dmitriev}, \citenamefont {Krasnikhin}, \citenamefont {Mulders},
  \citenamefont {Senin}, \citenamefont {Volovik},\ and\ \citenamefont
  {Yudin}}]{Dmi.10}%
  \BibitemOpen
  \bibfield  {author} {\bibinfo {author} {\bibfnamefont {V.~V.}\ \bibnamefont
  {Dmitriev}}, \bibinfo {author} {\bibfnamefont {D.~A.}\ \bibnamefont
  {Krasnikhin}}, \bibinfo {author} {\bibfnamefont {N.}~\bibnamefont {Mulders}},
  \bibinfo {author} {\bibfnamefont {A.~A.}\ \bibnamefont {Senin}}, \bibinfo
  {author} {\bibfnamefont {G.~E.}\ \bibnamefont {Volovik}}, \ and\ \bibinfo
  {author} {\bibfnamefont {A.~N.}\ \bibnamefont {Yudin}},\ }\href@noop {}
  {\bibfield  {journal} {\bibinfo  {journal} {JETP Lett.}\ }\textbf {\bibinfo
  {volume} {91}},\ \bibinfo {pages} {599} (\bibinfo {year} {2010})}\BibitemShut
  {NoStop}%
\bibitem [{\citenamefont {Li}\ \emph {et~al.}(2015)\citenamefont {Li},
  \citenamefont {Zimmerman}, \citenamefont {Pollanen}, \citenamefont
  {Collett},\ and\ \citenamefont {Halperin}}]{Li.15}%
  \BibitemOpen
  \bibfield  {author} {\bibinfo {author} {\bibfnamefont {J.~I.~A.}\
  \bibnamefont {Li}}, \bibinfo {author} {\bibfnamefont {A.~M.}\ \bibnamefont
  {Zimmerman}}, \bibinfo {author} {\bibfnamefont {J.}~\bibnamefont {Pollanen}},
  \bibinfo {author} {\bibfnamefont {C.~A.}\ \bibnamefont {Collett}}, \ and\
  \bibinfo {author} {\bibfnamefont {W.~P.}\ \bibnamefont {Halperin}},\ }\href
  {\doibase 10.1103/PhysRevLett.114.105302} {\bibfield  {journal} {\bibinfo
  {journal} {Phys. Rev. Lett.}\ }\textbf {\bibinfo {volume} {114}},\ \bibinfo
  {pages} {105302} (\bibinfo {year} {2015})}\BibitemShut {NoStop}%
\bibitem [{\citenamefont {{Halperin}}(2018)}]{Hal.18}%
  \BibitemOpen
  \bibfield  {author} {\bibinfo {author} {\bibfnamefont {W.~P.}\ \bibnamefont
  {{Halperin}}},\ }\href@noop {} {\bibfield  {journal} {\bibinfo  {journal}
  {ArXiv e-prints}\ } (\bibinfo {year} {2018})},\ \Eprint
  {http://arxiv.org/abs/1806.06437} {arXiv:1806.06437 [cond-mat.supr-con]}
  \BibitemShut {NoStop}%
\bibitem [{\citenamefont {Hakonen\emph{ et al.}}(1989)}]{Hak.89}%
  \BibitemOpen
  \bibfield  {author} {\bibinfo {author} {\bibfnamefont {P.~J.}\ \bibnamefont
  {Hakonen\emph{ et al.}}},\ }\href@noop {} {\bibfield  {journal} {\bibinfo
  {journal} {J. Low Temp. Phys.}\ }\textbf {\bibinfo {volume} {76}},\ \bibinfo
  {pages} {225} (\bibinfo {year} {1989})}\BibitemShut {NoStop}%
\bibitem [{\citenamefont {Li}\ \emph {et~al.}(2014{\natexlab{a}})\citenamefont
  {Li}, \citenamefont {Zimmerman}, \citenamefont {Pollanen}, \citenamefont
  {Collett}, \citenamefont {Gannon},\ and\ \citenamefont {Halperin}}]{Li.14b}%
  \BibitemOpen
  \bibfield  {author} {\bibinfo {author} {\bibfnamefont {J.~I.~A.}\
  \bibnamefont {Li}}, \bibinfo {author} {\bibfnamefont {A.~M.}\ \bibnamefont
  {Zimmerman}}, \bibinfo {author} {\bibfnamefont {J.}~\bibnamefont {Pollanen}},
  \bibinfo {author} {\bibfnamefont {C.~A.}\ \bibnamefont {Collett}}, \bibinfo
  {author} {\bibfnamefont {W.~J.}\ \bibnamefont {Gannon}}, \ and\ \bibinfo
  {author} {\bibfnamefont {W.~P.}\ \bibnamefont {Halperin}},\ }\href {\doibase
  10.1103/PhysRevLett.112.115303} {\bibfield  {journal} {\bibinfo  {journal}
  {Phys. Rev. Lett.}\ }\textbf {\bibinfo {volume} {112}},\ \bibinfo {pages}
  {115303} (\bibinfo {year} {2014}{\natexlab{a}})}\BibitemShut {NoStop}%
\bibitem [{\citenamefont {Li}\ \emph {et~al.}(2014{\natexlab{b}})\citenamefont
  {Li}, \citenamefont {Zimmerman}, \citenamefont {Pollanen}, \citenamefont
  {Collett}, \citenamefont {Gannon},\ and\ \citenamefont {Halperin}}]{Li.14a}%
  \BibitemOpen
  \bibfield  {author} {\bibinfo {author} {\bibfnamefont {J.~I.~A.}\
  \bibnamefont {Li}}, \bibinfo {author} {\bibfnamefont {A.~M.}\ \bibnamefont
  {Zimmerman}}, \bibinfo {author} {\bibfnamefont {J.}~\bibnamefont {Pollanen}},
  \bibinfo {author} {\bibfnamefont {C.~A.}\ \bibnamefont {Collett}}, \bibinfo
  {author} {\bibfnamefont {W.~J.}\ \bibnamefont {Gannon}}, \ and\ \bibinfo
  {author} {\bibfnamefont {W.~P.}\ \bibnamefont {Halperin}},\ }\href {\doibase
  10.1007/s10909-013-0917-3} {\bibfield  {journal} {\bibinfo  {journal} {J. Low
  Temp. Phys.}\ }\textbf {\bibinfo {volume} {175}},\ \bibinfo {pages} {31}
  (\bibinfo {year} {2014}{\natexlab{b}})}\BibitemShut {NoStop}%
\bibitem [{\citenamefont {Collin}\ \emph {et~al.}(2009)\citenamefont {Collin},
  \citenamefont {Triqueneaux}, \citenamefont {Bunkov},\ and\ \citenamefont
  {Godfrin}}]{Col.09}%
  \BibitemOpen
  \bibfield  {author} {\bibinfo {author} {\bibfnamefont {E.}~\bibnamefont
  {Collin}}, \bibinfo {author} {\bibfnamefont {S.}~\bibnamefont {Triqueneaux}},
  \bibinfo {author} {\bibfnamefont {Y.~M.}\ \bibnamefont {Bunkov}}, \ and\
  \bibinfo {author} {\bibfnamefont {H.}~\bibnamefont {Godfrin}},\ }\href@noop
  {} {\bibfield  {journal} {\bibinfo  {journal} {Phys. Rev. B}\ }\textbf
  {\bibinfo {volume} {80}},\ \bibinfo {pages} {094422} (\bibinfo {year}
  {2009})}\BibitemShut {NoStop}%
\bibitem [{\citenamefont {Li}\ \emph {et~al.}(2013)\citenamefont {Li},
  \citenamefont {Pollanen}, \citenamefont {Zimmerman}, \citenamefont {Collett},
  \citenamefont {Gannon},\ and\ \citenamefont {Halperin}}]{Li.13}%
  \BibitemOpen
  \bibfield  {author} {\bibinfo {author} {\bibfnamefont {J.~I.~A.}\
  \bibnamefont {Li}}, \bibinfo {author} {\bibfnamefont {J.}~\bibnamefont
  {Pollanen}}, \bibinfo {author} {\bibfnamefont {A.~M.}\ \bibnamefont
  {Zimmerman}}, \bibinfo {author} {\bibfnamefont {C.~A.}\ \bibnamefont
  {Collett}}, \bibinfo {author} {\bibfnamefont {W.~J.}\ \bibnamefont {Gannon}},
  \ and\ \bibinfo {author} {\bibfnamefont {W.~P.}\ \bibnamefont {Halperin}},\
  }\href@noop {} {\bibfield  {journal} {\bibinfo  {journal} {Nature Physics}\
  }\textbf {\bibinfo {volume} {9}},\ \bibinfo {pages} {775} (\bibinfo {year}
  {2013})}\BibitemShut {NoStop}%
\bibitem [{\citenamefont {Leggett}(1973)}]{Leg.73}%
  \BibitemOpen
  \bibfield  {author} {\bibinfo {author} {\bibfnamefont {A.~J.}\ \bibnamefont
  {Leggett}},\ }\href {http://stacks.iop.org/0022-3719/6/i=21/a=023} {\bibfield
   {journal} {\bibinfo  {journal} {Journal of Physics C: Solid State Physics}\
  }\textbf {\bibinfo {volume} {6}},\ \bibinfo {pages} {3187} (\bibinfo {year}
  {1973})}\BibitemShut {NoStop}%
\bibitem [{\citenamefont {Zimmerman}\ \emph {et~al.}()\citenamefont
  {Zimmerman}, \citenamefont {Nguyen},\ and\ \citenamefont
  {Halperin}}]{Zim.18}%
  \BibitemOpen
  \bibfield  {author} {\bibinfo {author} {\bibfnamefont {A.~M.}\ \bibnamefont
  {Zimmerman}}, \bibinfo {author} {\bibfnamefont {M.}~\bibnamefont {Nguyen}}, \
  and\ \bibinfo {author} {\bibfnamefont {W.~P.}\ \bibnamefont {Halperin}},\
  }\href@noop {} {\bibinfo  {journal} {submitted to J. Low Temp. Phys. (2018),
  arXiv:1808.07943 cond-mat.supr-con}\ }\BibitemShut {NoStop}%
\bibitem [{\citenamefont {Brinkman}\ and\ \citenamefont
  {Smith}(1975)}]{Bri.75}%
  \BibitemOpen
\bibfield  {journal} {  }\bibfield  {author} {\bibinfo {author} {\bibfnamefont
  {W.~F.}\ \bibnamefont {Brinkman}}\ and\ \bibinfo {author} {\bibfnamefont
  {H.}~\bibnamefont {Smith}},\ }\href@noop {} {\bibfield  {journal} {\bibinfo
  {journal} {Phys. Lett.}\ }\textbf {\bibinfo {volume} {51}},\ \bibinfo {pages}
  {449} (\bibinfo {year} {1975})}\BibitemShut {NoStop}%
\bibitem [{\citenamefont {Corruccini}\ and\ \citenamefont
  {Osheroff}(1978)}]{Cor.78}%
  \BibitemOpen
  \bibfield  {author} {\bibinfo {author} {\bibfnamefont {L.~R.}\ \bibnamefont
  {Corruccini}}\ and\ \bibinfo {author} {\bibfnamefont {D.~D.}\ \bibnamefont
  {Osheroff}},\ }\href {\doibase 10.1103/PhysRevB.17.126} {\bibfield  {journal}
  {\bibinfo  {journal} {Phys. Rev. B}\ }\textbf {\bibinfo {volume} {17}},\
  \bibinfo {pages} {126} (\bibinfo {year} {1978})}\BibitemShut {NoStop}%
\bibitem [{\citenamefont {Bunkov}\ and\ \citenamefont
  {Volovik}(1993)}]{Yur.93b}%
  \BibitemOpen
  \bibfield  {author} {\bibinfo {author} {\bibfnamefont {Y.~M.}\ \bibnamefont
  {Bunkov}}\ and\ \bibinfo {author} {\bibfnamefont {G.~E.}\ \bibnamefont
  {Volovik}},\ }\href@noop {} {\bibfield  {journal} {\bibinfo  {journal}
  {Europhys. Lett.}\ }\textbf {\bibinfo {volume} {21}},\ \bibinfo {pages} {837}
  (\bibinfo {year} {1993})}\BibitemShut {NoStop}%
\bibitem [{\citenamefont {Dmitriev\emph{ et al.}}(1999)}]{Dmi.99}%
  \BibitemOpen
  \bibfield  {author} {\bibinfo {author} {\bibfnamefont {V.~V.}\ \bibnamefont
  {Dmitriev\emph{ et al.}}},\ }\href@noop {} {\bibfield  {journal} {\bibinfo
  {journal} {Phys. Rev. B}\ }\textbf {\bibinfo {volume} {59}},\ \bibinfo
  {pages} {165} (\bibinfo {year} {1999})}\BibitemShut {NoStop}%
\bibitem [{\citenamefont {de~Gennes}\ and\ \citenamefont
  {Prost}(1995)}]{deG.95}%
  \BibitemOpen
  \bibfield  {author} {\bibinfo {author} {\bibfnamefont {P.~G.}\ \bibnamefont
  {de~Gennes}}\ and\ \bibinfo {author} {\bibfnamefont {J.}~\bibnamefont
  {Prost}},\ }\href@noop {} {\emph {\bibinfo {title} {The Physics of Liquid
  Crystals}}}\ (\bibinfo  {publisher} {Oxford University Press},\ \bibinfo
  {address} {Great Clarendon Street, Oxford OX2 6DP, UK},\ \bibinfo {year}
  {1995})\BibitemShut {NoStop}%
\bibitem [{\citenamefont {Jongh}\ and\ \citenamefont {Miedema}(2001)}]{DeJ.01}%
  \BibitemOpen
  \bibfield  {author} {\bibinfo {author} {\bibfnamefont {L.~J.~D.}\
  \bibnamefont {Jongh}}\ and\ \bibinfo {author} {\bibfnamefont {A.~R.}\
  \bibnamefont {Miedema}},\ }\href {\doibase 10.1080/00018730110101412}
  {\bibfield  {journal} {\bibinfo  {journal} {Advances in Physics}\ }\textbf
  {\bibinfo {volume} {50}},\ \bibinfo {pages} {947} (\bibinfo {year} {2001})},\
  \Eprint {http://arxiv.org/abs/https://doi.org/10.1080/00018730110101412}
  {https://doi.org/10.1080/00018730110101412} \BibitemShut {NoStop}%
\bibitem [{\citenamefont {Song}\ \emph {et~al.}(2009)\citenamefont {Song},
  \citenamefont {Chung}, \citenamefont {Park},\ and\ \citenamefont
  {Choi}}]{Son.09}%
  \BibitemOpen
  \bibfield  {author} {\bibinfo {author} {\bibfnamefont {Y.-S.}\ \bibnamefont
  {Song}}, \bibinfo {author} {\bibfnamefont {J.-H.}\ \bibnamefont {Chung}},
  \bibinfo {author} {\bibfnamefont {J.~M.~S.}\ \bibnamefont {Park}}, \ and\
  \bibinfo {author} {\bibfnamefont {Y.-N.}\ \bibnamefont {Choi}},\ }\href
  {\doibase 10.1103/PhysRevB.79.224415} {\bibfield  {journal} {\bibinfo
  {journal} {Phys. Rev. B}\ }\textbf {\bibinfo {volume} {79}},\ \bibinfo
  {pages} {224415} (\bibinfo {year} {2009})}\BibitemShut {NoStop}%
\bibitem [{\citenamefont {Yamashita}\ \emph {et~al.}(2008)\citenamefont
  {Yamashita}, \citenamefont {Izumina}, \citenamefont {Matsubara},
  \citenamefont {Sasaki}, \citenamefont {Ishikawa}, \citenamefont {Takagi},
  \citenamefont {Kubota},\ and\ \citenamefont {Mizusaki}}]{Yam.08}%
  \BibitemOpen
  \bibfield  {author} {\bibinfo {author} {\bibfnamefont {M.}~\bibnamefont
  {Yamashita}}, \bibinfo {author} {\bibfnamefont {K.}~\bibnamefont {Izumina}},
  \bibinfo {author} {\bibfnamefont {A.}~\bibnamefont {Matsubara}}, \bibinfo
  {author} {\bibfnamefont {Y.}~\bibnamefont {Sasaki}}, \bibinfo {author}
  {\bibfnamefont {O.}~\bibnamefont {Ishikawa}}, \bibinfo {author}
  {\bibfnamefont {T.}~\bibnamefont {Takagi}}, \bibinfo {author} {\bibfnamefont
  {M.}~\bibnamefont {Kubota}}, \ and\ \bibinfo {author} {\bibfnamefont
  {T.}~\bibnamefont {Mizusaki}},\ }\href {\doibase
  10.1103/PhysRevLett.101.025302} {\bibfield  {journal} {\bibinfo  {journal}
  {Phys. Rev. Lett.}\ }\textbf {\bibinfo {volume} {101}},\ \bibinfo {pages}
  {025302} (\bibinfo {year} {2008})}\BibitemShut {NoStop}%
\bibitem [{\citenamefont {Halperin}\ \emph {et~al.}(2008)\citenamefont
  {Halperin}, \citenamefont {Choi}, \citenamefont {Davis},\ and\ \citenamefont
  {Pollanen}}]{Hal.08}%
  \BibitemOpen
  \bibfield  {author} {\bibinfo {author} {\bibfnamefont {W.~P.}\ \bibnamefont
  {Halperin}}, \bibinfo {author} {\bibfnamefont {H.}~\bibnamefont {Choi}},
  \bibinfo {author} {\bibfnamefont {J.~P.}\ \bibnamefont {Davis}}, \ and\
  \bibinfo {author} {\bibfnamefont {J.}~\bibnamefont {Pollanen}},\ }\href@noop
  {} {\bibfield  {journal} {\bibinfo  {journal} {J. Phys. Soc. Jpn.}\ }\textbf
  {\bibinfo {volume} {77}},\ \bibinfo {pages} {111002} (\bibinfo {year}
  {2008})}\BibitemShut {NoStop}%
\bibitem [{\citenamefont {Abrikosov}\ and\ \citenamefont
  {Gorkov}(1962)}]{Abr.62}%
  \BibitemOpen
  \bibfield  {author} {\bibinfo {author} {\bibfnamefont {A.~A.}\ \bibnamefont
  {Abrikosov}}\ and\ \bibinfo {author} {\bibfnamefont {L.~P.}\ \bibnamefont
  {Gorkov}},\ }\href@noop {} {\bibfield  {journal} {\bibinfo  {journal} {J.
  Exp. Theor. Phys.}\ }\textbf {\bibinfo {volume} {15}},\ \bibinfo {pages}
  {752–757} (\bibinfo {year} {1962})}\BibitemShut {NoStop}%
\bibitem [{\citenamefont {Larkin}(1965)}]{Lar.65}%
  \BibitemOpen
  \bibfield  {author} {\bibinfo {author} {\bibfnamefont {A.~I.}\ \bibnamefont
  {Larkin}},\ }\href@noop {} {\bibfield  {journal} {\bibinfo  {journal} {Sov.
  Phys. JETP}\ }\textbf {\bibinfo {volume} {2}},\ \bibinfo {pages} {130}
  (\bibinfo {year} {1965})}\BibitemShut {NoStop}%
\bibitem [{\citenamefont {Fomin}(2008)}]{Fom.08}%
  \BibitemOpen
  \bibfield  {author} {\bibinfo {author} {\bibfnamefont {I.~A.}\ \bibnamefont
  {Fomin}},\ }\href {\doibase 10.1134/S0021364008130134} {\bibfield  {journal}
  {\bibinfo  {journal} {JETP Letters}\ }\textbf {\bibinfo {volume} {88}},\
  \bibinfo {pages} {59} (\bibinfo {year} {2008})}\BibitemShut {NoStop}%
\bibitem [{\citenamefont {Sauls}\ and\ \citenamefont {Sharma}(2003)}]{Sau.03}%
  \BibitemOpen
  \bibfield  {author} {\bibinfo {author} {\bibfnamefont {J.~A.}\ \bibnamefont
  {Sauls}}\ and\ \bibinfo {author} {\bibfnamefont {P.}~\bibnamefont {Sharma}},\
  }\href@noop {} {\bibfield  {journal} {\bibinfo  {journal} {Phys. Rev. B}\
  }\textbf {\bibinfo {volume} {68}},\ \bibinfo {pages} {224502} (\bibinfo
  {year} {2003})}\BibitemShut {NoStop}%
\end{thebibliography}
\end{document}